\documentstyle[prb,aps,epsfig,floats,amsfonts]{revtex}

\begin{document}

\twocolumn
\psfull
\draft

\title{Resistivity of a Metal between the Boltzmann Transport Regime and the Anderson Transition}
\author{Branislav K. Nikoli\' c and Philip B. Allen}
\address{Department of Physics and Astronomy,
SUNY at Stony Brook, Stony Brook, New York 11794-3800}

\maketitle

\begin{abstract}
We study the transport properties of a finite three dimensional
disordered conductor, for both weak and strong scattering on
impurities, employing the real-space Green function technique and
related Landauer-type formula. The dirty metal is described by a
nearest neighbor tight-binding Hamiltonian with a single s-orbital
per site and random on-site potential (Anderson model). We compute
exactly the zero-temperature conductance of a finite-size sample
placed between two semi-infinite disorder-free leads. The
resistivity is found from the coefficient of linear scaling of the
disorder-averaged resistance with sample length. This ``quantum''
resistivity is compared to the semiclassical Boltzmann expression
computed in both Born approximation and multiple scattering
approximation.
\end{abstract}



Ever since Anderson's seminal paper,~\cite{anderson} a prime model
for the theories of the disorder induced metal-insulator, or
localization-delocalization~\cite{janssen} (LD), transition in
non-interacting electron systems has been the tight-binding
Hamiltonian (TBH) on a hypercubic lattice
\begin{equation}\label{eq:tbh}
  \hat{H} = \sum_{\bf m} \varepsilon_{\bf m}|{\bf m} \rangle \langle {\bf m}|
  +  t\sum_{\langle {\bf m},{\bf n} \rangle}
  |{\bf m} \rangle \langle {\bf n}|,
\end{equation}
with nearest neighbor hopping matrix element $t$ between
s-orbitals $\langle {\bf r}|{\bf m} \rangle = \psi({\bf r}-{\bf
m})$ on adjacent atoms located at sites ${\bf m}$ of the lattice.
The disorder is simulated by taking random on-site potential such
that $\varepsilon_{\bf m}$ is uniformly distributed in the
interval [-W/2,W/2]. This is commonly called the ``Anderson
model''. There are many numerical studies~\cite{kramer} of the LD
transition, which occurs in three-dimensions (3D) for a
half-filled band at the critical disorder strength~\cite{slevin}
$W_c \approx 16.5t$. Experiments on real metals with strong
scattering or strong correlations often yield resistivities which
are hard to analyze. Theory gives guidance in two extreme
regimes: (a) the semiclassical case where quasiparticles with
definite ${\bf k}$ vector justify a Boltzmann approach and ``weak
localization'' (WL) correction,~\cite{wl} and (b) a scaling
regime~\cite{gang4} near the LD transition to ``strong
localization''. Lacking a complete theory it is often assumed
that the two limits join smoothly with nothing between.
Experiments, however, are very often in neither extreme limit.
The middle is wide and needs more attention.

\begin{figure}
\centerline{\psfig{file=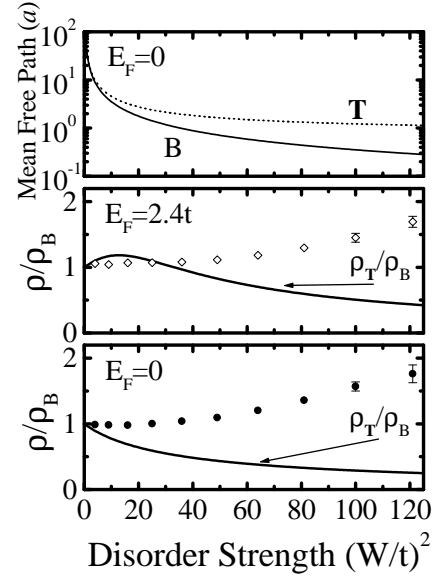,height=3.0in,angle=0} }
\vspace{0.2in}
  \caption{Resistivity $\rho$ at $E_F=0$ (lower
panel) and $E_F=2.4t$ (middle panel), from a sample of cross
section $A=225 \, a^2$, normalized to the semiclassical Boltzmann
resistivity $\rho_{\text{B}}$ calculated in the Born
approximation. Also plotted are the ratios of $\rho_{\text {B}}$
to the Boltzmann resistivity $\rho_{\bf T}$ obtained using a
${\bf T}$-matrix for multiple scattering on a single impurity.
The upper panel shows putative mean free paths obtained from
$\rho_{\text{B}}$ (labeled by B) or $\rho_{\bf T}$ (labeled by
{\bf T}). Error bars at small $W/t$ are smaller than the size of
the dot.}
 \label{fig:rho}
\end{figure}

Here we give a 3D numerical analysis focused not on the
transition itself but instead on the resistivity for
$1<W/t<W_c/t$; specifically we ask how rapidly does the
resistivity $\rho(W)$ deviate from the values predicted by the
usual Boltzmann theory valid when $W \ll t$. It has long been
assumed that ``Ioffe-Regel condition''~\cite{yoffe} $\ell \sim 1/k_F
\sim a$ ($\ell$ being the mean free path, and $a$ being the lattice
constant) gives the criterion for sufficient disorder to drive
the metal into an Anderson insulator. Figure~\ref{fig:rho} shows
that this is wrong.  By $W/t \sim 4$,
where $\ell$ is close to $a$, there is little sign of a
divergence away from the semiclassical extrapolation, and
the LD transition is postponed to much larger values of $W/t$.
A cleaner discussion is possible using Kubo theory, which does not
define $\ell$, but allows a definition of the diffusivity $D_i$ of
an eigenstate $|i \rangle$, as shown below in
Eq.~(\ref{eq:statediff}). In the semiclassical regime, $D_i
\rightarrow D_k=v_k \ell_k/3$. The diffusivity $D_k$ diminishes as
$(W/t)^{-2}$ in Boltzmann theory. As $\ell/a$ approaches a minimum
value ($\sim 1$), $D_i$ decreases toward
$D_{\text{min}}=ta^2/\hbar$, which can be regarded as a minimum
metallic diffusivity below which localization sets in. But there
is a wide range of $W/t$ over which $D_i \lesssim D_{\text{min}}$
and yet the Boltzmann scaling $D \sim (t/W)^2$ is approximately
right. In this regime single particle eigenstates $|i \rangle$ are
neither ballistically propagating nor are they localized. There is
a third category: intrinsically diffusive.~\cite{allen} A wave
packet built from such states has zero range of ballistic motion
but an infinite range of diffusive propagation. Such states are
found not only in a narrow crossover regime but over a wide range
of parameters physically accessible in real materials and
mathematically accessible in models like the Anderson model. In
this regime, there is not a simple scaling parameter nor a
universal behavior. But the behavior is quite insensitive to a
changes in Fermi energy $E_F$ or $k_B T$, and scales smoothly with
$W/t$.

The traditional tool for computation of $\rho$ has been the Kubo
formula,~\cite{kubo} originally derived for a system in the
thermodynamic limit. In a basis of exact single particle electron
state $|i\rangle$ of energy $\epsilon_i$, this can be written as
\begin{equation}\label{eq:kubo-difusivity}
    \sigma=\frac{1}{\rho}=\frac{e^2}{\Omega} \sum_i \left
    (-\frac{\partial f}{\partial \epsilon_i} \right )
    D_i=e^2 N(E_F)\bar{D},
\end{equation}
where $\Omega$ is the sample volume, $f$ is the equilibrium
Fermi-Dirac distribution, $N(E_F)$ the density of states at $E_F$,
$\bar{D}$ the mean diffusivity, and state diffusivity is given by
\begin{equation}\label{eq:statediff}
  D_i=\pi \hbar \sum_j |\langle i|\hat{ v}_x|j \rangle|^2
\delta(\epsilon_i-\epsilon_j)
\end{equation}
where $\hat{\bf v}$ is the velocity operator. These formulas,
while correct, are hard to use numerically.~\cite{nikolic0} Thanks
to the recent advances in mesoscopic physics,~\cite{datta} it is
now apparent that the Landauer scattering
approach~\cite{landauer} (or equivalent ``mesoscopic'' Kubo
reformulation for the finite-size systems~\cite{verges}) provides
superior numerical efficiency when computing the transport
properties of finite disordered conductors.
It relates the conductance of a sample to its quantum-mechanical
transmission properties. This formalism emphasizes the importance
of taking into account the interfaces between the sample and the
rest of the circuit.~\cite{landimry} Transport in the sample is
phase-coherent (i.e. effectively occurring at zero temperature);
the dissipation and thus thermalization of electrons (necessary
for the establishment of steady state) takes place in other parts
of the circuit.

Our principal result for the (quantum) resistivity of Anderson
model, using Landauer-type approach, is shown on
Fig.~\ref{fig:rho} for two different Fermi energies $E_F=0$
(half-filled band) and $E_F=2.4t$ (approximately 70\% filled band
but falling somewhat as $W$, and thus the band-width, increases).
The linearized Boltzmann equation $-e {\bf E} \cdot {\bf v}_{\bf
k} \partial f/\partial \epsilon_{\bf k} = (dF_{\bf
k}/dt)_{\text{scatt}}$ serves as a reference theory. Here
$\epsilon_{\bf k}$ is the energy band for $W=0$, namely
$\epsilon_{\bf k}=2t \sum \cos k_{\alpha}$, $\hbar v_{k \alpha}$
is $\partial \epsilon/\partial k_{\alpha}$, and $F_{\bf k}$ is
the non-equilibrium distribution. The collision integral is
\begin{equation}\label{eq:integral}
  \left(  \frac{dF}{dt} \right)_{\text{scatt}} = -\frac{2
  \pi}{\hbar}
  \sum_{{\bf k}^{\prime}} |V_{{\bf k k}^{\prime}}|^2 (F_{\bf
  k}-F_{{\bf k}^{\prime}}) \delta(\epsilon_{\bf k}-\epsilon_{\bf
  k'}).
\end{equation}
The mean squared matrix element of the random potential $|V_{{\bf
k k}^{\prime}}|^2$, in Born approximation, is
$\overline{\varepsilon^2_{\bf m}}=W^2/12$, where
$\overline{(\ldots)}$  denotes average over probability
distribution $P(\varepsilon_{\bf m})=(1/W)
\theta(W/2-|\varepsilon_{\bf m}|)$. This equation assumes that
quasiparticles propagate with mean free path $\ell > a$ between
isolated collision events. The equation is exactly solvable,
yielding (for $k_B T \ll t$) $1/\rho_{\text B}=e^2
\tau(n/m)_{\text{eff}}$, with $(n/m)_{\text{eff}}=\sum v_{kx}^2
\delta(\epsilon_k-E_F)/\Omega$, and $\hbar/\tau=2\pi N(E_F)
W^2/12$. We have evaluated $(n/m)_{\text{eff}}$ and $N(E_F)$
numerically. To within factors of order one, the Boltzmann-Born
answer for the semiclassical resistivity is $\rho_{\text B}=(\pi
\hbar a/e^2)(W/4t)^2$. When $W=3t$ and $a=3 {\text \AA}$,
$\rho_{\text B}$ is $125\, \mu \Omega$cm, typical of dirty
transition metal alloys, and close to the largest resistivity
normally seen in dirty ``good'' metals. Figure~\ref{fig:rho} plots
$\rho/\rho_{\text B}$ versus $(W/t)^2$. Even for $W=10t$ there is
less than a factor of 2 deviation from the (unwarranted)
extrapolation of the Boltzmann theory into the regime $W > t$.
Boltzmann theory can be ``improved'' by including multiple
scattering from single impurities, that is, replacing the
impurity potential by the ${\bf T}$-matrix ${\bf T}_{\bf m}
(z)=\varepsilon_{\bf m}/(1-\varepsilon_{\bf m}g(z))$ where
$g(z)=(1/N_s) \sum (z-\epsilon_{\bf k})^{-1}$ is the free
particle Green function ($N_s$ is the number of lattice sites).
To next order the mean square ${\bf T}$-matrix is
\begin{equation}\label{eq:tmatrix}
  \overline{|{\bf T}_{\bf m}(z)|^2}=\frac{W^2}{12} \left (1+ \frac{3W^2}{20t^2}( g
  g^*+gg+g^*g^*) + \ldots \right),
\end{equation}
where the first term is the Born approximation and the coefficient
of the correction ($\sim {\mathcal O}(W^4)$) changes sign from
negative to positive as $E_F$ moves from 0 to $2.4t$. As shown on
Fig.~\ref{fig:rho}, the resistivity does not behave like
$\overline{|{\bf T}_{\bf m}(z)|^2}$; multiple scattering with
interference from pairs of impurities is at least equally
important, and the ``exact'' $\rho(W)$ is less sensitive to
details like $E_F$ than is the ${\bf T}$-matrix approximation. The
rest of the paper presents the method of calculation and
describes a bit of mesoscopic physics of very dirty metals.

\begin{figure}
\centerline{\psfig{file=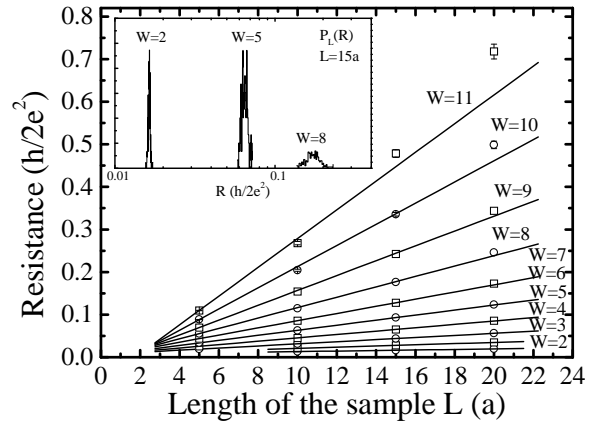,height=3.0in,angle=-90} }
\vspace{0.2in} \caption{Linear fit $\langle R \rangle=C_1+\rho \,
L/A$, ($A=225 \,a^2$) for the disorder-averaged resistance
$\langle R \rangle$ in the band center $E_F=0$ and different
disorder strengths $W$. The intercept $C_1$ is decreasing with
increasing $W$ (i.e. it is not determined just by the contact
resistance $\pi \hbar/147e^2$) and becomes negative for $W
\protect \gtrsim 7t$. The inset shows examples of the
distribution of resistances $P_L(R)$ (for $L=15a$) versus $\log
R$. The distribution broadens either by increasing $W$ or the
length of the sample (the units on the $y$-axis are arbitrary and
different for each distribution).} \label{fig:linearfit}
\end{figure}

The central linear transport quantity in the mesoscopic
view,~\cite{janssen} as well as in the scaling theory of
localization,~\cite{gang4} is conductance $G$ rather than
conductivity $\sigma(L) = L^{2-d}G(L)$ (the bulk conductivity is
an intensive material constant defined only in the thermodynamic
limit, $\sigma=\lim_{L \rightarrow \infty} L^{2-d}G(L)$). We use a
Landauer-type formula to get the exact quantum conductance $G$ of
finite samples with disorder configurations chosen by a random
number generator. Finite-size samples permit exact solutions for any
strength of disorder. Similar to other recent
works,~\cite{kahnt,todorov} the bulk resistivity is extracted from
the disorder-averaged resistance $\langle R \rangle$ by finding
the linear (Ohmic) scaling of $\langle R \rangle$ versus the
length of the sample $L$ at fixed cross section $A$
(Fig.~\ref{fig:linearfit}). Two kinds of errors~\cite{todorov} may
arise: (a) The transition from the Ohmic regime to the localized
regime occurs for length of the sample $L \sim \xi$ which happens
when~\cite{thouless} $G = {\mathcal O}(2e^2/h)$. If $L$ is made
large enough, $G$ will always diminish to this magnitude.
Therefore, we avoid using the sample sizes with too small $G$.
(b) Finite-size boundary conditions and non-specular
reflection~\cite{sondh} cause density of states and scattering
properties of the sample to be slightly altered as compared to
the true bulk. We expect these effects to be small for our
samples where $\ell$ is smaller than the transverse size
$\sqrt{A}$.

A two probe measuring configuration is used for computation. The
sample is placed between two disorder-free ($\varepsilon_{\bf
m}=0$) semi-infinite leads connected to macroscopic reservoirs
which inject thermalized electrons at electrochemical potential
$\mu_L$ (from the left) or $\mu_R$ (from the right) into the
system. The electrochemical potential difference $e
V=\mu_L-\mu_R$ is measured between the reservoirs. The leads have
the same cross section as the sample. The hopping parameter in
the lead and the one which couples the lead to the sample are
equal to the hopping parameter in the sample. Thus, extra
scattering (and resistance) at the sample-lead interface is
avoided but transport at Fermi energies $|E_F|$ greater than the
clean-metal band edge $|E_b|=6t$ cannot be studied.~\cite{nikolic0}
Hard wall boundary conditions are used in the $\hat{y}$ and $\hat{z}$
directions. The sample is modeled on a cubic lattice with $N
\times N_y \times N_z$ sites, where $N_y=N_z=15$ and lengths
$L=Na$ are taken from the set $N \in \{5,10,15,20\}$.

The linear conductance is calculated using an expression obtained
from the Keldysh technique~\cite{caroli}
\begin{equation}\label{eq:greenlandauer}
  G  =  \frac{4e^2}{\pi \hbar} \, \text{Tr} \left ( \text{Im} \, \hat{\Sigma}_L \, \hat{G}^{r}_{1 N} \,
  \text{Im} \, \hat{\Sigma}_R \, \hat{G}^{a}_{N 1} \right ).
\end{equation}
Here $\text{Im} \,
\hat{\Sigma}_{L,R}=(\hat{\Sigma}_{L,R}^r-\hat{\Sigma}_{L,R}^a)/2i$
are self-energy matrices ($r$-retarded, $a$-advanced) which
describe the coupling of the sample to the leads, and
$\hat{G}^{r}_{1 N}$, $\hat{G}^{a}_{N 1}$ are Green function
matrices connecting the layer $1$ and $N$ of the sample:
$\hat{G}^{r,a}=(E-\hat{H}-\hat{\Sigma}^{r,a})^{-1}$
($\hat{G}^{a}=[\hat{G}^{r}]^{\dagger}$), with
$\hat{\Sigma}^{r}=\hat{\Sigma}_L^{r}+\hat{\Sigma}_R^{r}$
($\hat{\Sigma}^{a}=[\hat{\Sigma}^{r}]^{\dagger})$. The self-energy
matrices introduced by the leads are non-zero only on the end
layers of the sample adjacent to the leads. They are
given~\cite{datta} by $\hat{\Sigma}^{r}_{L,R}({\bf n},{\bf m}) =
t^2 \hat{g}_{L,R}^r({\bf n}_S,{\bf m}_S)$ with
$\hat{g}_{L,R}^r({\bf n}_S,{\bf m}_S)$ being the surface Green
function~\cite{verges} of the bare semi-infinite lead between the
sites ${\bf n}_S$ and ${\bf m}_S$ in the end atomic layer of the
lead (adjacent to the corresponding sites ${\bf n}$ and ${\bf m}$
inside the conductor). Positive definiteness of the operators
$-2\, \text{Im} \, \hat{\Sigma}_{L,R}$ makes it possible to find
their square root and recast the expression under the trace of
Eq.~(\ref{eq:greenlandauer}) as a Hermitian operator. The
expression~(\ref{eq:greenlandauer}) then looks like the Landauer
formula involving the transmission matrix ${\bf t}$
\begin{eqnarray}\label{eq:ttlandauer}
  G & = & \frac{e^2}{\pi \hbar} \, \text{Tr} \, ({\bf t t}^{\dag}) =
  \frac{e^2}{\pi \hbar} \sum_{n=1}^{N_y N_z} T_n, \\
  {\bf t} & = & 2 \sqrt{-\text{Im} \, \hat{\Sigma}_L} \, \hat{G}^{r}_{1 N}
  \sqrt{-\text{Im}\, \hat{\Sigma}_R},
  \label{eq:t}
\end{eqnarray}
or transmission eigenvalues $T_n$ when the trace is evaluated in a
basis which diagonalizes ${\bf t t}^{\dag}$.

For the case of two probe geometry the average transmission in the
semiclassical transport regime ($a < \ell \ll L \ll \xi $) is
given by~\cite{datta} $\langle T \rangle = \ell_0/(\ell_0+L)$,
with $\ell_0$ being of the order of $\ell$.
Thus, the semiclassical limit~\cite{semilandauer} of the Landauer
formula for conductance $\langle G \rangle = (e^2/\pi \hbar)\, M
\langle T \rangle$ (measured between points deep inside the
reservoirs) in the case of not too strong scattering should have
the form
\begin{equation}\label{eq:semilandauer}
  \langle G \rangle^{-1} =R_C+ \rho \frac{L}{A}.
\end{equation}
It describes the (classical) series addition of two resistors. The
``contact'' resistance~\cite{sharvin} $R_C=\pi \hbar/e^2 M$ is
non-zero, even in the case of ballistic transport when the second
term containing the resistivity $\rho=(\pi \hbar/e^2) \, A/\ell_0
M$ vanishes. Here $M \sim k_F^2 A$ is the number of propagating
transverse modes at $E_F$, also referred to as ``channels''. A
ballistic conductor with a finite cross section can carry only
finite currents (the voltage drop occurs at the lead-reservoir
interface). Using this simple analysis for guidance, we plot
average resistances (taken over $N_{\text{conf}}=200$ realization
of disorder) versus $L$ in Fig.~\ref{fig:linearfit}, and fit with
the linear function
\begin{equation} \label{eq:fit}
\langle R \rangle =C_1+C_2L, \ C_2=\rho/A.
\end{equation}
The resistivity $\rho$ on Fig.~\ref{fig:rho} is obtained from the
fitted value of $C_2$. For very small values of $W$ the constant $C_1$
is approximately equal to $R_C=\pi \hbar/e^2 M$ (where $M=147$ is
the number of open channels in the band center). To
our surprise, $C_1$ diminishes steadily with increasing $W$, and
even turns negative around $W \gtrsim 7t$.

\begin{figure}
\centerline{\psfig{file=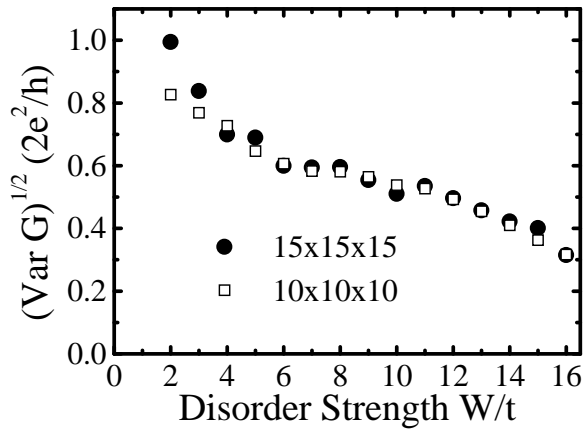,height=3.0in,angle=-90} }
\vspace{0.2in} \caption{The conductance fluctuations ($\Delta G
=\sqrt{\text{Var} \, G}$ at $E_F=0$) from weak to strong
scattering regime in the disordered cubic samples $10 \times 10
\times 10$ and $15 \times 15 \times 15$.} \label{fig:ucf}
\end{figure}

The quantum conductance $G$ fluctuates from sample to sample
exhibiting universal conductance fluctuations~\cite{ucf} (UCF)
$\Delta G =\sqrt{\text{Var} \, G} \simeq e^2/\pi \hbar$ in the
semiclassical transport regime $G \gg e^2/\pi \hbar$. The inset on
Fig.~\ref{fig:linearfit} shows the distribution of
resistance~\cite{cohen} $P_L(R)$ for our numerically generated
impurity ensemble. The error bars, used as weights in the
fit~(\ref{eq:fit}), are computed as $\delta \langle R \rangle =
\sqrt{\text{Var} R/N_{\text{conf}}}$. We find that $\Delta G$ is
indeed independent of the size $L$ (of cubic samples), but
decreases systematically by a factor $\approx 3$ as $W$ increases
to the critical value $W_c$ (Fig.~\ref{fig:ucf}). On the other
hand, $\Delta R$, being similar to $\Delta G/G^2$, depends on
sample size. As $W$ approaches $W_c$, $G$ gets smaller until (for
our finite samples) $\Delta G/G \sim 1$. At this point the
distribution of resistances $R=1/G$ becomes very broad and
$\langle R \rangle$ begins to rise above $1/\langle G \rangle$.
For $L=15$ this happens when $W \gtrsim 12t$. At large $W$ the
conductance of long samples ($N=20$) becomes close to $e^2/\pi
\hbar$ and deviations from Ohmic scaling are expected. Therefore,
we do not use these points in the fitting procedure when $W
\gtrsim 10t$ (keeping the conductance~\cite{todorov} of the
fitted samples $G > 2e^2/\pi \hbar$).

We do not have a complete explanation for the deviation of
$C_1$~(\ref{eq:fit}) from the quantum point contact resistance
$R_C$. In the semiclassical regime $G \gg e^2/\pi \hbar$ there are
corrections to the Ohmic scaling $G \propto L^{d-2}$. The
Diffuson-Cooperon diagrammatic perturbation theory gives a
(negative) WL correction~\cite{wl}
\begin{equation}\label{eq:wl}
  \sigma(L)=\sigma + \frac{e^2}{\pi^2 \hbar 2 \sqrt{2}} \frac{1}{L}-
  \frac{e^2}{\pi^3 \hbar}\frac{1}{\ell_0^{\prime}},
\end{equation}
where $\ell_0^{\prime}$ is a length of order $\ell$ (its precise
value does not lead to observable consequences in the experiments
studying WL, as long as it is unaffected by the temperature and
the magnetic field). The positive $1/L$ term in Eq.~\ref{eq:wl}
provides a possible picture for our finding that $C_1$in
Eq.~(\ref{eq:fit}) goes negative as $W$ increases. However, this
picture is an extrapolation from the semiclassical into the
``middle'' regime of intrinsically diffusive states, and
therefore should be given little weight. The negative values of
$C_1$ is better regarded as a new numerical result from the
mesoscopic dirty metal theory.

It is interesting to note that in many $d$-band intermetallic
compounds, $\rho$ ``saturates'' at a constant value~\cite{allen2}
rather than following the semiclassical extrapolation, that is,
increasing linearly with $T$ at high $T$.  High $T_c$ materials
and doped C$_{60}$ metals, on  the other hand, do not
saturate.~\cite{allen2} Within Boltzmann theory, the static
disorder measured by $(W/t)^2$ plays the same role as thermal
disorder or squared lattice displacement $\propto k_B T$. Our
numerical results thus can be described as ``failing to
saturate.''  Similar failure was seen in high $T$ Monte Carlo
studies by Gunnarsson and Han.~\cite{gunnarsson}

We thank I. L. Aleiner for interesting discussions and challenging
questions. Suggestions provided by J. A. Verg\' es are
acknowledged. This work was supported in part by NSF grant no. DMR
9725037.

\end{document}